\input harvmac
\sequentialequations
\overfullrule=0pt
\def\bigZ{Z\!\!\!Z}
\def\uA{\,\lower 1.2ex\hbox{$\sim$}\mkern-13.5mu A}
\def\sw{{\scriptscriptstyle\rm SW}}
\def\oneloop{\buildrel{\scriptscriptstyle\rm 1\hbox{-}loop}\over =}
\def\DRED{{\scriptscriptstyle\rm DRED}}
\def\DREG{{\scriptscriptstyle\rm DREG}}
\def\zero{{\scriptscriptstyle(0)}}
\font\authorfont=cmcsc10 \ifx\answ\bigans\else scaled\magstep1\fi
{\divide\baselineskip by 4
\multiply\baselineskip by 4
\def\prenomat{\matrix{\hbox{hep-th/9611016}&\cr \qquad
  \hbox{DO-LE/96}  
&\cr}}
\Title{$\prenomat$}{\vbox{\centerline{On $N=2$
Supersymmetric QCD with 4 Flavors
}}}
\centerline{\authorfont Nicholas Dorey}
\bigskip
\centerline{\sl Physics Department, University College of Swansea}
\centerline{\sl Swansea SA2$\,$8PP UK $\quad$ \tt n.dorey@swansea.ac.uk}
\bigskip
\centerline{\authorfont Valentin V. Khoze}
\bigskip
\centerline{\sl Department of Physics, Centre for Particle Theory, 
University of Durham}
\centerline{\sl Durham DH1$\,$3LE UK $\quad$ \tt valya.khoze@durham.ac.uk}
\bigskip
\centerline{and}
\bigskip
\centerline{\authorfont Michael P. Mattis}
\bigskip
\centerline{\sl Theoretical Division T-8, Los Alamos National Laboratory}
\centerline{\sl Los Alamos, NM 87545 USA$\quad$ \tt mattis@pion.lanl.gov}
\vskip .3in
\def\hf{{\textstyle{1\over2}}}
\def\quarter{{\textstyle{1\over4}}}
\noindent
Seiberg and Witten's proposed solution of $N=2$ SQCD with $N_c=2$ and
$N_F=4$ is known
 to conflict with instanton calculations in three distinct ways.
Here we show how to resolve all three discrepancies, 
simply by reparametrizing
the elliptic curve in terms of quantities 
$\tau^\zero_{\rm eff}$ and $\tilde{u}$ 
rather than  $\tau$ and $u = \langle\Tr\uA^2\rangle$.
 $SL(2,\bigZ)$ invariance of the
curve is preserved. However,
there is now an infinite ambiguity in the relation between 
 $\tau^\zero_{\rm eff}$ and $\tau$ and between $\tilde{u}$ and $u$, 
corresponding
to an infinite number of unknown coefficients in the instanton expansion.
Thus the reinterpreted curve (unlike the cases $N_F<4$)
no longer determines the quantum modulus
$u$ as a function of the classical VEV $a$.
\vskip .1in
\Date{\bf November 1996 } 
\vfil\break
}

\lref\Siegel{W. Siegel, Phys.~Lett.~84B (1979) 193.}
\lref\ISletter{K. Ito and N. Sasakura, \it One-instanton calculations
in $N=2$ $SU(N_c)$ Supersymmetric QCD\rm, 
 hep-th/9609104.}
\lref\Fucito{F. Fucito and G. Travaglini, 
{\it Instanton calculus and nonperturbative relations
 in $N=2$ supersymmetric gauge theories}, 
 ROM2F-96-32, hep-th/9605215.}
\lref\SVone{M.A. Shifman and A.I. Vainshtein, 
Nucl. Phys. B277 (1986) 456, Nucl. Phys. B359 (1991) 571.}
\lref\Koblitz{See for example N. Koblitz, {\it Introduction To Elliptic 
Curves And Modular Forms} (Springer-Verlag, 1984), Sec.~I.6.}
\lref\Dine{M. Dine and Yu. Shirman, Phys. Rev. D50 (1994) 5389.}
\lref\Cel{W. Celmaster and R. Gonsalves, Phys. Rev. D20 (1979) 1420}
\lref\Hone{L. Hall, Nucl. Phys. B178 (1981) 75.}
\lref\Wone{S. Weinberg, Phys. Lett. 91B (1980) 51.}
\lref\MVone{S.P. Martin, M.T. Vaughn, Phys. Lett. B318 (1993) 331.}
\lref\DHone{E. D'Hoker, I. M. Krichever and D. H. Phong,
 {\it The effective prepotential of $N=2$ supersymmetric
  $SU(N)$ gauge theories}, hep-th/9609041}
\lref\dkmone{N. Dorey, V.V. Khoze and M.P. Mattis, \it Multi-instanton
calculus in $N=2$ supersymmetric gauge theory\rm, hep-th/9603136,
Phys.~Rev.~D54 (1996) 2921.}
\lref\dkmtwo{N. Dorey, V.V. Khoze and M.P. Mattis, 
\it Multi-instanton check of the relation between the prepotential  
${\cal F}$ 
and the modulus $u$ in $N=2$ SUSY Yang-Mills theory\rm,
hep-th/9606199, Phys.~Lett.~B389 (in press, 26 December 1996).}
\lref\dkmthree{N. Dorey, V.V. Khoze and M.P. Mattis, 
\it A two-instanton test of the exact solution of $N=2$ supersymmetric  
QCD\rm,
hep-th/9607066, Phys.~Lett.~B388 (1996) 324.}
\lref\dkmfour{N. Dorey, V.V. Khoze and M.P. Mattis, \it Multi-instanton
calculus in $N=2$ supersymmetric gauge theory.
II. Coupling to matter\rm, hep-th/9607202,
Phys.~Rev.~D (in press, 15 December 1996).}
\lref\HS{T. Harano and M. Sato, \it Multi-instanton
calculus versus exact results in N=2 supersymmetric QCD\rm,
hep-th/9608060}
\lref\Aoyama{H. Aoyama, T. Harano, M. Sato and S.Wada,
\it   Multi-instanton calculus in N=2 supersymmetric QCD\rm,
hep-th/9607076.}   
\lref\SWone{N. Seiberg and E. Witten, 
{\it Electric-magnetic duality, monopole
condensation, and confinement in $N=2$ supersymmetric Yang-Mills  
theory}, 
Nucl. Phys. B426 (1994) 19, (E) B430 (1994) 485  hep-th/9407087}
\lref\SWtwo{
N. Seiberg and E. Witten, 
{\it Monopoles, duality and chiral symmetry breaking
in $N=2$ supersymmetric QCD}, 
Nucl. Phys B431 (1994) 484 ,  hep-th/9408099}
\lref\West{P.S. Howe and P. West, 
{\it Superconformal Ward identities and $N=2$ Yang-Mills theory},
hep-th/9607239.}
\lref\BMSTY{T. Eguchi and S.K. Yang, 
{\it Prepotentials of $N=2$ supersymmetric gauge theories 
and soliton equations}, hep-th/9510183;
\hfil\break
G. Bonelli and M. Matone, 
    Phys. Rev. Lett. 76 (1996) 4107, hep-th/9602174; \hfil\break
 J. Sonnenschein, S. Theisen and S. Yankielowicz,
 Phys. Lett. 367B (1996) 145, hep-th/9510129.}
\lref\BMT{ G. Bonelli, M. Matone and M. Tonin, 
{\it Solving $N=2$ SYM by the reflection symmetry of quantum vacua},
hep-th/9610026.}
\lref\MAtone{
M. Matone, {\it Instantons and recursion relations in $N=2$ SUSY gauge  
theory}
 Phys. Lett. B357 (1995) 342,    hep-th/9506102.  }
\lref\FPone{ D. Finnell and P. Pouliot,
{\it Instanton calculations versus exact results in 4 dimensional 
SUSY gauge theories},
Nucl. Phys. B453 (95) 225, hep-th/9503115. }
\lref\Sone{ N. Seiberg, Phys. Lett. B206 (1988) 75. }
\lref\tHooft{G. 't Hooft, Phys. Rev. D14 (1976) 3432; ibid.
D18 (1978) 2199.}
\def\frac#1#2{{ {#1}\over{#2}}}

\def\eff{{\rm eff}}
\def\qeff{q_\eff}

\def\sst{\scriptscriptstyle}
\def\dreg{\sst\rm DREG}
\def\dred{\sst\rm DRED}
\def\pv{\sst\rm PV}
\def\PV{{\sst\rm PV}}
\def\Mw{M_{ W}}

\def\F{{\cal F}}

\def\cl{{\,\rm cl}}
\def\lambdabar{\bar\lambda}

\def\psibar{\bar\psi}

\def\tL{\Lambda^{\scriptscriptstyle\rm SW}}

\def\vhiggs{{\rm v}}
\def\vhiggsd{{\rm v}_D}

\def\infinity{\infty}

\def\new{{\scriptscriptstyle\rm new}}

\def\uX{\,\lower 1.2ex\hbox{$\sim$}\mkern-13.5mu X}
\def\uD{\,\lower 1.2ex\hbox{$\sim$}\mkern-13.5mu {\rm D}}

\def\uF{\,\lower 1.2ex\hbox{$\sim$}\mkern-13.5mu F}
\def\uW{\,\lower 1.2ex\hbox{$\sim$}\mkern-13.5mu W}
\def\uWbar{\,\lower 1.2ex\hbox{$\sim$}\mkern-13.5mu {\overline W}}
\def\uPhibar{\,\lower 1.2ex\hbox{$\sim$}\mkern-13.5mu {\overline \Phi}}

\def\uV{\,\lower 1.2ex\hbox{$\sim$}\mkern-13.5mu V}
\def\uv{\,\lower 1.0ex\hbox{$\scriptstyle\sim$}\mkern-11.0mu v}
\def\uPsi{\,\lower 1.2ex\hbox{$\sim$}\mkern-13.5mu \Psi}
\def\uPhi{\,\lower 1.2ex\hbox{$\sim$}\mkern-13.5mu \Phi}
\def\uchi{\,\lower 1.5ex\hbox{$\sim$}\mkern-13.5mu \chi}
\def\Psibar{\bar\Psi}
\def\uPsibar{\,\lower 1.2ex\hbox{$\sim$}\mkern-13.5mu \Psibar}
\def\upsi{\,\lower 1.5ex\hbox{$\sim$}\mkern-13.5mu \psi}
\def\psibar{\bar\psi}
\def\upsibar{\,\lower 1.5ex\hbox{$\sim$}\mkern-13.5mu \psibar}
\def\upsibarzero{\,\lower 1.5ex\hbox{$\sim$}\mkern-13.5mu \psibar^\zero}
\def\ulambda{\,\lower 1.2ex\hbox{$\sim$}\mkern-13.5mu \lambda}
\def\ulambdabar{\,\lower 1.2ex\hbox{$\sim$}\mkern-13.5mu \lambdabar}
\def\ulambdabarzero{\,\lower 1.2ex\hbox{$\sim$}\mkern-13.5mu \lambdabar^\zero}
\def\ulambdabarnew{\,\lower 1.2ex\hbox{$\sim$}\mkern-13.5mu \lambdabar^\new}
\def\D{{\cal D}}

\def\Dslash{\,\,{\raise.15ex\hbox{/}\mkern-12mu \D}}
\def\Dbarslash{\,\,{\raise.15ex\hbox{/}\mkern-12mu {\bar\D}}}
\def\delslash{\,\,{\raise.15ex\hbox{/}\mkern-9mu \partial}}
\def\delbarslash{\,\,{\raise.15ex\hbox{/}\mkern-9mu {\bar\partial}}}
\def\L{{\cal L}}
\def\hf{{\textstyle{1\over2}}}
\def\quarter{{\textstyle{1\over4}}}
\def\eighth{{\textstyle{1\over8}}}

\def\uAcl{\,\lower 1.2ex\hbox{$\sim$}\mkern-13.5mu A^{}_{\cl}}
\def\uAbarcl{\,\lower 1.2ex\hbox{$\sim$}\mkern-13.5mu A_{\cl}^\dagger}

\def\Leff{\L_{\rm eff}}

\def\geff{g_{\rm eff}}

\def\taueff{\tau_{\rm eff}}
\def\thetaeff{\theta_{\rm eff}}

\newsec{Review of discrepancies between the Seiberg-Witten solutions
and instanton calculations}

The solutions of $N=2$ supersymmetric QCD with gauge group $SU(2)$ 
proposed by Seiberg and Witten \SWtwo\ have recently been tested against 
first-principles instanton calculations, up to the two-instanton
level \refs{\dkmthree\dkmfour-\Aoyama}. As reviewed below,
these tests have resulted in perfect agreement when $N_F,$ the number of
quark hypermultiplet flavors, is $\le2.$ On the other hand,
interesting discrepancies have emerged for the cases $N_F=3$ \Aoyama\ and
$N_F=4$ \dkmfour. In  this paper, we present our conjecture for
how to resolve these discrepancies for $N_F=4.$ Just like the
conjectured resolution for $N_F=3$ \HS, we will not actually need to modify
the elliptic curve that governs the solution (Eq.~(16.35) of \SWtwo).
Instead, we will change the interpretation of the parameters $\tau$ and
$\tilde u$ that appear in this expression; they will have a different
(and highly underdetermined)
relation to physical observables of the microscopic theory
than that proposed by Seiberg and Witten. In particular, in sharp 
contrast to the cases $N_F<4,$ the reinterpreted
curve for $N_F=4$ no longer determines the quantum modulus
\eqn\udef{u\ =\ \langle\Tr\uA^2\rangle \ , }
as a function of the classical VEV $a$ of the adjoint Higgs  $\uA =
A^a \tau^a/2$.
  
The solutions given in \SWtwo\ make precise predictions for all
multi-instanton contributions to the low-energy physics.
The functional form of these contributions is tightly constrained by 
holomorphy, the $U(1)_{R}$ anomaly, renormalization group (RG) invariance,
 and the  Matone relation \refs{\MAtone\BMSTY-\BMT} between the prepotential
$\F(a)$ and the  modulus $u(a)$.
 For $N_{F}>0$, additional constraints come from flavor symmetries, 
the decoupling limit for heavy flavors, and a discrete $\bigZ_2$
symmetry  forbidding odd-instanton contributions when any quark mass
vanishes.
As we showed in Refs.~\refs{\dkmthree,\dkmfour,\dkmone,\dkmtwo},
 all these constraints are
 built into the  instanton calculus as well.\foot{The 
all-instanton-orders proof of the Matone relation given in \dkmtwo\
was ostensibly limited to the case $N_F=0.$ However, the reader can
easily verify that the proof goes through for arbitrary numbers of (massless
or massive) hypermultiplets. The key point here is that Eqs. (20)-(21)
of \dkmtwo\ can be directly extended to $N_F >0$, see Eq. (7.20)
of \dkmfour. 
The Matone relation can also be understood as a Ward identity
for (broken) superconformal invariance \West.}
Therefore, in testing the Seiberg-Witten solutions against explicit
instanton calculations, one can  ``mod out'' these constraints,
and focus on a  basic subset of nontrivial numerical predictions extracted
from the elliptic curves at each order in the instanton expansion.
These numbers are then compared against the results of a finite-dimensional
integration over the (multi-)instanton super-moduli.
  
\vskip .3in
\vbox{
\halign{\hfil\bf#&\quad\hfil#\hfil&\quad\hfil#\hfil&\quad#\hfil\cr
{}&\sl parameter &\sl first nontrivial prediction&\sl
first nontrivial prediction \cr
$N_F$&\sl fixing& \sl of curve ($\,\ge1$ massless quark)
&\sl of curve (all $m_q>0\,$)\cr
{}&{}&{}&{}\cr
0&1-loop&1-inst&N/A\cr
1&1-loop&2-inst&2-inst\cr
2&1-loop&2-inst&2-inst\cr
3&1-loop+2-inst&4-inst&3-inst\cr
{}&[Eq.~(4)]&{}&{}\cr
4&1-loop&\it none \rm if all $m_q=0$;&3-inst\cr
{}&$+\ $all-even-insts&otherwise 4-inst&{}\cr{}&[Eqs.~(6), (42)]&{}&{}\cr
}
\vskip .1in
{\narrower\smallskip
\noindent\bf Table 1\rm. For each $N_F\le4,$ the second column gives the
order in the semiclassical expansion at which the input parameters
$\tL_{N_F}$ and $\tilde u$ (or $\taueff^\zero$ and
$\tilde u$ for $N_F=4$) in the Seiberg-Witten curves are fixed
in terms of physical quantities in the microscopic theory. The third
and fourth columns (respectively, at least one massless quark, and all
quarks massive) then give the  order in the semiclassical
expansion at which a nontrivial numerical prediction is first made
by the  curve, which can be compared against an 
instanton calculation (``nontrivial'' is defined in the text).
 To date, no numerical
tests beyond the 2-instanton level have been performed, so that the
models with $N_F=0,1,2$ are on a firmer footing than the ones
with $N_F=3,4$. 
\smallskip}
}
\vskip .2in

In formulating these tests, and evaluating their outcome,
it is crucial to distinguish between
what is \it input \rm (i.e., the fixing of conventions) versus
what is \it output \rm (i.e., a definite numerical prediction).
Since this distinction will be key to resolving the discrepancies for $N_F=4,$
we should first review the models with $N_F\le3\,$; see Table 1  for
a summary. The input parameters
for the elliptic curves with $0\le N_F\le3$ are the dynamically generated scale
$\tL_{N_F}$ and the modulus $\tilde u$, as well as the $N_F$ quark masses. 
(Notation: henceforth we will denote by $\tilde u$ the parameter in the 
Seiberg-Witten curves,
and reserve the symbol $u$ always to mean  $\langle\Tr\uA^2\rangle$.)
The curve defines an implicit choice of regularization scheme for
$\tL_{N_F}$. For any physical
quantity the $n$-instanton contributions
are proportional to $(\tL_{N_F})^{n(4-N_F)}\ $; 
in order to test the specific numerical predictions of the curves 
against instantons, one needs to know
how to relate $\tL_{N_F}$ to (say) the analogous quantity $\Lambda_{N_F}^\PV$
in the Pauli-Villars  (PV)
scheme which is most natural for instanton calculations
\tHooft. As explained by Finnell and Pouliot \FPone\ and reviewed
in Sec.~2 below, this ``dictionary'' is fixed at the 1-loop level, and reads:
\eqn\relrel{ (\tL_{N_F})^{4-N_F} \ = \ 4 (\Lambda^\PV_{N_F})^{4-N_F} \ ,
\qquad N_F=0,1,2,3\ . } 
Similarly, for $0\le N_F\le2,$ 
a discrete symmetry on the $u$ plane combined with
a semiclassical analysis of the
singularities gives simply \SWtwo\
\eqn\SWsimple{\tilde u\ =\ u\ ,\qquad N_F=0,1,2\ .}
Equations~\relrel-\SWsimple\ are the input; the predictions for $\F(a)$
and $u(a)$ first
lie at the 1-instanton level for $N_F=0$, and at the 2-instanton level
for $N_F=1,2$ due to the $\bigZ_2$ symmetry mentioned above.\foot{When
quark masses are nonzero the models with $N_F=1,2$ have
1-instanton contributions too; however, due to built-in
 heavy-quark decoupling properties, these contributions are
pegged to the 1-instanton term
for $N_F=0$ and so are not independent tests \refs{\dkmthree,\dkmfour}.}
The tests for $N_F=0$ have been performed at the 1-instanton 
\FPone\ and at the 2-instanton \dkmone\ level, and the tests for
$N_F=1,2$ have been carried out at the 2-instanton level 
\refs{\dkmthree-\Aoyama}, all with exact agreement.

Next we consider the case  $N_F=3.$ Here Seiberg and Witten
(in Sec.~14.1 of \SWtwo) continue to assume the equality \SWsimple,
even though it is no longer required by
a symmetry argument. Indeed, 
for $N_F=3,$ the $\bigZ_{4-N_F}$ symmetry that acts on the complex
$u$ plane is trivial \SWtwo; consequently Eq.~\SWsimple\ can
be generalized to $\tilde u= u-u_0\cdot
(\Lambda^{\scriptscriptstyle\rm SW}_3)^2$ where $u_0$ is a numerical
constant.
(This possibility is noted by Seiberg and Witten, but not exploited.)
A nonzero value for $u_0$ has been extracted from a 2-instanton calculation by
Aoyama et al; they find \Aoyama 
\eqn\unought{\tilde u\ =\  u-u_0\cdot
(\Lambda^{\scriptscriptstyle\rm SW}_3)^2\ ,\quad u_0\ =\ -{1\over2^43^3}\ ,
\qquad N_F=3}
contradicting the naive assumption \SWsimple. It is natural to
hypothesize \HS\ that the solution to the $N_F=3$ model is still given by
the Seiberg-Witten curve (Eq.~(14.5) of \SWtwo), with the substitution
\unought\ rather than \SWsimple. 
In sum, for $N_F=3,$ both the 1-loop level and the 2-instanton level should
be considered as \it input \rm due to the extra degree of freedom
$u_0$; the first testable predictions extracted from the curve then
lie at the 3-instanton level (or, if one of the quarks is
massless, at the 4-instanton level).

\def\LSW{\Lambda^{\scriptscriptstyle\rm SW}}
\def\LSWNF{\Lambda^{\scriptscriptstyle\rm SW}_{N_F}}
In this paper we will focus on the particularly interesting case $N_F=4.$ 
Here, unlike the previous cases, the proposed solution and
the instanton calculation disagree at the level of
 the low-energy effective $U(1)$ Lagrangian $\Leff$ \dkmfour. 
In fact, 
this disagreement can already be seen in the massless model. In this case
both the $\beta$-function and the $U(1)_R$ anomaly vanish identically, so that
the microscopic coupling $g_4$ and $\theta$-parameter $\theta_4$ 
(which can no longer be rotated away) assemble
into a single scale-independent holomorphic coupling 
$\tau   =  {8 \pi i \over g_4^2 } +    {\theta_4  \over  \pi}\,.$
 Furthermore $\Leff^\zero$  
is simply the Lagrangian  of a classical $N=2$ free 
field theory; its overall normalization, $\taueff^\zero,$ enters the
BPS formulae for the dyon masses \SWtwo. (The superscript
$\scriptstyle(0)$ will denote  the massless case.) 
In Section 10 of \SWtwo\ Seiberg and
Witten make the strong additional assumption that, thanks to the
absence of a running coupling constant, the effective $U(1)$ coupling
 equals the classical $SU(2)$ coupling,
\eqn\exactness{\taueff^\zero\ = \ \tau\ ,}
which implies no quantum (perturbative or non-perturbative)
corrections to $\tau$.
Instead, as we demonstrated in Ref.~\dkmfour, a first-principles instanton
calculation gives\foot{Note that we use
$q$ for the 1-instanton factor rather than for the 2-instanton factor
as in \SWtwo.}
\eqn\tauresult{
\taueff^\zero\ \equiv\ \hf\F^{\zero\prime\prime}(a)\  =\ 
\tau\ +\ {i\over \pi}\sum_{n=0,2,4\ldots}\,
c_n\,q^{n}\ ,\qquad q\ \equiv \ \exp(i\pi\tau) \ .}
In particular a nonzero 2-instanton contribution $c_2=-{7\over2^6\,3^5}$
 was calculated in Ref.~\dkmfour.
 The 1-loop perturbative constant $c_0$, while not considered in
Ref.~\dkmfour,  will turn out to be crucial to our proposed resolution.
We  calculate below that  $c_0=4\log 2$ in the PV
scheme.\foot{If the classical exactness assumption \exactness\  were correct,
there would be no need to introduce a regularization scheme, since
 scheme dependence is a  one-loop effect.
As discussed in Sec.~3, although the $N_F=4$ model is a finite theory, 
the microscopic $SU(2)$ 
coupling $g_4$ must still be defined in a particular scheme;
that is because the finiteness is due to cancellations between individually
UV divergent graphs.}

A second  disagreement in the $N_F=4$ model
arises when at least one of the four hypermultiplets
has a nonzero mass, say $m_4\neq0$. In the double scaling
limit defined by $m_4\rightarrow\infty$ together with  $g_4\rightarrow0$
in a specific way reviewed below, the heavy flavor decouples, and the
model is supposed to flow to the $N_F=3$ theory. This requires 
the identification of the dynamical scale $\LSWNF$ for $N_F=3,$ with
the parameters of the $N_F=4$ theory. 
As explained below, working in the PV scheme 
and using the dictionary
\relrel, one obtains:
\eqn\PVSW{\tL_{3} \ = \ 4 \Lambda^\PV_{3} \ = \
4m_4\,\exp(-8\pi^2/(g^\PV_{4})^2)\ ,}
 where $g^\PV_4$ is the microscopic PV coupling in the
4-flavor model.  In contrast, the  relation 
given in \SWtwo,
\eqn\SWSW{\tL_{3} \ =  \
64m_4\,\exp(-8\pi^2/(g_4^\sw)^2)\ ,}
involves a proportionality constant of 64  rather than
4. Again, if the quantum corrections to Eq.~\exactness\
were absent, there could be no distinction
between the PV coupling constant $g^\PV_4$
and the ``classical" $g_4^\sw$ of Seiberg and Witten, hence no
accounting for the factor of 16 mismatch between 
Eqs.~\PVSW\ and \SWSW.

Thirdly, even with the identification \SWSW, the specific
$N_F=4$ solution proposed in \SWtwo\ flows  to the uncorrected version
of the $N_F=3$ model which fails to incorporate the shifted definition
of $\tilde u$ given in Eq.~\unought.

Our principal aim is to explain how to resolve all three of
these $N_F=4$ discrepancies in a simple way, 
through a reinterpretation of the quantities $\tau$ and $\tilde u$
that enter into the massive Seiberg-Witten curve. In particular, rather
than being modular forms of the microscopic $SU(2)$ parameter $\tau,$ 
 the coefficients of the massive curve  will be functions of the
effective \it massless \rm $U(1)$ coupling $\taueff^\zero$
defined by the all-even-instanton series \tauresult.
(Obviously the factor of 16 between
Eqs.~\PVSW\ and \SWSW\ will be automatically accounted for by
the exponentiation of $c_0=4\log 2$.)
This redefinition preserves the important $SL(2,\bigZ)$ invariance
of the elliptic curve, as well as the stringent residue condition
described in Sec.~17 of \SWtwo. However,
it also introduces an infinite ambiguity into
the solution, parametrized by the
infinite number of as-yet-undetermined numerical coefficients $c_4,$
$c_6,$ $\ldots,$ in Eq.~\tauresult\ (a similar series, Eq.~(42),
relates $\tilde u$ and $u$). These issues are discussed in
Sec.~3 below, which is devoted to the case $N_F=4$, and also contains
our conclusions.  But first, in Sec.~2, we
lay some necessary groundwork in the cases $N_F<4.$ In particular we
review Weinberg's matching prescription between high- and low-energy
gauge theories \Wone, as this formalism lies at the heart of the physics.
In the process we will also specify the perturbative,
one-loop contributions to the
prepotential for $N_F<4,$ correcting some incomplete expressions
in the literature. 

The main result of this paper is the reinterpreted 
$N_{F}=4$ massive  curve which agrees with all available 
perturbative and multi-instanton calculations. Although the input
parameters of this curve receive contributions from all even numbers
of instantons, the curve does contain definite numerical predictions
at the 3-instanton level (assuming nonvanishing quark masses) that
can, in principle, be tested against a semiclassical calculation.
We anticipate that similar reinterpretations need to be made in the
general class of $N=2$ models with gauge group $SU(N_c)$ and $N_F=2N_c$
for which the $\beta$-function vanishes. (Indeed, discrepancies at the
1-instanton level have  been claimed for $N_c=3,$ $N_F=4,6$
by Ito and Sasakura \ISletter.)


\newsec{Weinberg's matching prescription and the Seiberg-Witten
regularization scheme for $N_F<4$}

The physics of $N=2$ SQCD utilizes---in two distinct but equally important
ways---a matching prescription between a ``high-energy'' and a ``low-energy''
gauge theory. On the one hand, the $SU(2)$ gauge group spontaneously breaks
down to $U(1)$ as the adjoint Higgs $\uA$ acquires a complex VEV 
$ \vev{\uA}=a \tau^3.$ For
energy scales $E\ll M_W=\sqrt{8}|a|,$ the dynamics is governed by
a nonrenormalizable Wilsonian effective action with $U(1)$ gauge invariance,
formally obtained by integrating out the heavy quanta. On the other
hand, one also needs to understand the RG decoupling mentioned earlier,
whereby the $SU(2)$ theory with $N_F$ flavors of quark hypermultiplets
flows to
the $SU(2)$ theory with $N_F-1$ flavors, in the limit that one of the
quarks becomes infinitely massive. 

\def\ghe{g_{\scriptscriptstyle\rm HE}}
\def\gle{g_{\scriptscriptstyle\rm LE}}
\def\bhe{b_{\scriptscriptstyle\rm HE}}
\def\ble{b_{\scriptscriptstyle\rm LE}}
\def\MSbar{\overline{\hbox{MS}}}
Both types of matching may be accomplished with the use of Weinberg's
one-loop \hbox{formula \Wone}
\eqn\Weone{ {1 \over \gle^2 (\mu)} \ \oneloop \ {1 \over \ghe^2 (\mu)}
- \lambda (\mu) \ .}
Here $\mu$ is the characteristic momentum scale of the light fields;
$\gle(\mu)$ and $\ghe(\mu)$ are renormalized 
gauge couplings of the low- and
high-energy theories, respectively;
and $\lambda (\mu)$ is a finite correction coming
from one-loop contributions of heavy particles to the gauge 
self-energy $\Sigma.$ Thus
$\ghe(\mu)$ is extracted from the complete set of
 one-loop contributions to $\Sigma,$ whereas for
$\gle(\mu)$ only light particles are permitted on the
  external and internal legs.
For example, in the $\MSbar$
scheme, $\ghe(\mu)$ and $\gle(\mu)$ are defined in $D$ dimensions in terms
of the bare couplings ${\ghe}_B$ and ${\gle}_B$
in the standard way:
\eqn\bare{{\ghe}_B \ \mu^{2-D/2} \ = \ \ghe(\mu) -
 \ \bhe \ \ghe^3(\mu) \  
\biggl({1 \over D-4} + \hf \gamma^{}_{\scriptscriptstyle\rm E}
-\hf \log 4\pi \biggr) \ , }
\eqn\barel{{\gle}_B \ \mu^{2-D/2} \ = \ \gle(\mu) -
 \ \ble \ \gle^3(\mu) \  
\biggl({1 \over D-4} + \hf \gamma^{}_{\scriptscriptstyle\rm E}
-\hf \log 4\pi \biggr) \ , }
where $\bhe$ and $\ble$ are the one-loop
coefficients of the corresponding $\beta$-functions.
Note that when the low-energy theory is
supersymmetric pure $U(1)$ gauge theory, $\gle$
receives no perturbative corrections and 
is scale-independent in the limit $D\rightarrow4$.

The quantity $\lambda$ has the generic  form
\eqn\lamgen{\lambda (\mu) \ = \ C_1 + C_2 \log {m_v \over \mu} 
+ C_3 \log {m_f \over \mu} + C_4 \log {m_s \over \mu} \ ,}
where $m_v$, $m_f$ and $m_s$ are the masses of the heavy
 vector,
fermion and scalar particles that have been integrated out.
The $C_i$ are group-theoretic constants that are
tabulated by Hall (see Appendix 1 of \Hone, in which an error in
\Wone\ is corrected). In particular $C_1$ and $C_2$ appear
only when the heavy vector particles have been integrated out;
similarly, $C_3$ appears when there are heavy fermions,
and $C_4$ corresponds to the heavy scalars.

The original calculations of \refs{\Wone-\Hone} were performed
in the dimensional regularization 
with $\MSbar$  scheme (DREG). However, they
can be easily translated into the supersymmetry-preserving 
dimensional reduction with $\MSbar$
scheme (DRED) \Siegel, or into the Pauli-Villars (PV) scheme.
The one-loop relations
between the coupling constants in different schemes 
for the $SU(N_{\rm c})$ gauge group
can be found for example in \refs{\MVone, \FPone}:
\eqn\schemes{{1 \over g_{\pv}^2 (\mu)} \ = \ 
{1 \over g_{\dred}^2 (\mu)} \ = \ 
{1 \over g_{\dreg}^2 (\mu)} + {1 \over 48 \pi^2} N_{\rm c} \ ,} 
independently of $N_F.$ 
It turns out that for the case of spontaneous 
symmetry breaking $[SU(2) , N_{F}]  \longrightarrow  [U(1),  0]$ 
considered in detail presently,
\eqn\pauli{   C_1^{\pv} \ = \ C_1^{\dred}\ =\ 0 \ . }
In contrast, in the case of the heavy flavor
decoupling $[SU(2),  N_{F}]  \longrightarrow$ $[SU(2),$ \hbox{$N_{F}-1]$},
vector particles do not decouple and $C_1 \equiv 0$ by definition.

Now let us apply this formalism, in turn, to these two cases of
interest.

\subsec{Spontaneous symmetry breaking}

We consider the Coulomb branch of $N=2$ SQCD, in which only the adjoint
Higgs $\uA\equiv A^a \tau^a/2$ acquires a VEV, 
say in the $\tau^3$ direction:
$\vev{\uA\,}=a\tau^3$. (We have adopted here the VEV normalization
conventions of \SWtwo; the translation
formulae to the original normalization of \SWone\
used in our previous work
are assembled in  Appendix A.) The $SU(2)$
component of the adjoint $N=2$ supermultiplet that is parallel to the
VEV then remains massless, while the components $\propto\ \tau^1$ or
$\tau^2$ acquire a mass $M_W=\sqrt{8}|a|.$ In addition there are $2N_F$
quark multiplets $Q_f$ and $\tilde Q_f,$ $f=1,\cdots,N_F,$ in the
fundamental representation of the gauge group. The `1' and `2' color
components of these multiplets acquire masses
$|\sqrt{2}a+m_f|$ and $|\sqrt{2}a-m_f|$, respectively,
 as can be seen from a tree-level examination of the
$N=2$ invariant superpotential \SWtwo
\eqn\superp{{\cal W}\ =\ \sum_{f=1}^{N_{F}} \sqrt{2}\tilde{Q}_{f}\uPhi  
Q_{f} +
m_{f}\tilde{Q}_{f}Q_{f}\ .}
Here $\uPhi$ is the $N=1$ adjoint
chiral superfield whose lowest component is $\uA$; color indices
are suppressed. 

In the PV or DRED schemes one then has
\eqn\lamsqcd{\lambda (\mu) \ = \ 
C_{\sst \rm adj} \log {\Mw \over \mu} 
\ +\  C_{\sst\rm fund} \sum_{f=1}^{N_{F}}\big(
\log {|\sqrt{2}a+m_f| \over \mu}  +
\log {|\sqrt{2}a-m_f| \over \mu}  \big)
\ ,}
where $C_{\sst \rm adj}$ and $C_{\sst \rm fund}$ are numerical constants.
Setting $\mu=M_W$ for simplicity and extracting
$C_{\sst \rm fund}  =  1/ 16 \pi^2$ from Ref.~\Hone, one finds
\eqn\newlamsqcd{\lambda (M_W) \ = \ 
{1\over16\pi^2} \sum_{f=1}^{N_{F}}
\log\, \left|{2a^2-m_f^2 \over 8a^2}\right|\ ,}
so that Eq.~\Weone\ becomes
\eqn\dkmmat{ 
{1 \over \geff^2}  \ \oneloop \ {4-N_{F} \over 8 \pi^2}
  \log\biggl({\Mw \over \Lambda_{N_{F}}}\biggr) \ + \ 
 {N_{F} \over 8 \pi^2}\log 2 \ - \ 
{1 \over 16 \pi^2} \sum_{f=1}^{N_{F}}
\ \log\,\big|1-m_f^2/2a^2\big|\ .}
In this expression
 $\geff\equiv \gle$ denotes the effective $U(1)$ coupling constant,
 $\Lambda_{N_F}$ is the PV or equivalently 
 the DRED dynamical scale (we drop the PV superscript henceforth)
 \eqn\LAMdef{ \Lambda_{N_F}^b \ =\
\mu^{b} \exp [-8\pi^2 /g_{\pv}^2 (\mu)]  \ =\
\mu^{b} \exp [-8\pi^2 /g_{\dred}^2 (\mu)] 
\ , }
and we have used the fact that the (negated) 
coefficient of the $\beta$-function
for these models is $b =  2 N_{ c} - N_F =  4-N_F $. We make the following
comments:

\bf(i) \rm
Equations \dkmmat-\LAMdef\ extend to $N_F\ge0$ the case of $N_F=0$ considered
by Finnell and Pouliot, who obtained simply \FPone
\eqn\FPmatch{ 
{1 \over \geff^2}  \ \oneloop \ {1 \over g_{\pv}^2 (\Mw)}\ =\ {1\over2\pi^2}
\log\biggl({M_W\over\Lambda_0}\biggr)\ .}
This is referred to as the absence of threshold corrections.

\bf(ii) \rm
As usual in a supersymmetric chiral theory \SVone, 
$1/\geff^2$ is the imaginary part of a holomorphic
complexified coupling constant
\eqn\tauequals{\taueff  \ = \ {8 \pi i \over \geff^2 } + 
   {\thetaeff  \over  \pi} \ .}
This implies that Eq.~\dkmmat\ may be analytically continued away from
the imaginary axis, as follows:
\eqn\newdkmmat{ \taueff
\ \oneloop \ (i/\pi)(4-N_{F})
  \log\biggl({\sqrt{8}a \over \Lambda_{N_{F}}}\biggr) \ + \ 
 {iN_{F} \over  \pi}\log 2 \ - \ 
{i \over 2 \pi} \sum_{f=1}^{N_{F}}
\ \log\,\big(\,1-m_f^2/2a^2\,\big)\ .}

\subsec{Heavy flavor decoupling}

Next we consider the case in which a single hypermultiplet flavor
becomes very heavy (say, $m_{N_F}\rightarrow\infty$) and decouples
from the spectrum, leaving behind $N=2$ SQCD with one fewer flavor.
Choosing $\mu=m_{N_F}$ we find that $\lambda(m_{N_F})=0$ as follows
from  Eq.~\lamgen\ with $C_1=C_2\equiv 0$; consequently 
\eqn\nothresh{g^{-2}_{N_F}(m_{N_F})\ =\ g^{-2}_{N_F-1}(m_{N_F})}
 in the PV or DRED schemes.
Rewriting this relation in an RG-invariant way using \LAMdef, one obtains
\eqn\Lamrecursion{m_{N_F}\cdot\Lambda_{N_F}^{4-N_F}\ =\ 
\Lambda_{N_F-1}^{4-(N_F-1)}\ .}
The appropriate double scaling limit is therefore defined by
$m_{N_F}\rightarrow\infty$ and $\Lambda_{N_F}\rightarrow0$ with
the product on the left-hand side of \Lamrecursion\ being held fixed.

In the remainder of this section we will relate the dynamical
scales $\LSW_{N_F}$ that appear in the Seiberg-Witten elliptic curves,
on the one hand,
 to the PV or DRED scales \LAMdef, on the other hand. This is done by
comparing Eqs.~\newdkmmat-\Lamrecursion\ to the explicit solutions
proposed in \SWtwo.

\subsec{Relating the Seiberg-Witten scheme to the PV or DRED schemes}

The Seiberg-Witten elliptic curve for the $N_{F}=0$ theory is \SWtwo\ 
\eqn\swzero{ y^2 \ = \ x^2(x-u) \ + \ {1 \over 4} (\tL_0)^4 x
 \ . }
The curve defines a dynamical scale $\tL_0$
in a particular scheme, the ``Seiberg-Witten scheme''; the superscript
SW
 is introduced to distinguish it from $\Lambda_0$
in the PV or DRED schemes.
The correspondence between the SW and the PV schemes for $N_F=0$
 has been examined by Finnell and Pouliot \FPone, using Weinberg's 
matching formula. From the elliptic curve \swzero, they calculate
$ \taueff  \simeq (2i/\pi)  \log\big(8 u / (\tL_0)^2\big)$
valid in the semiclassical regime, $u\simeq2a^2.$ Comparing this
result with Eq.~\newdkmmat\ with $N_F=0$ then yields \FPone
\eqn\pvsw{ (\tL_0)^4 \ = \ 4 \, \Lambda_0^4 \ .}
While this derivation was ostensibly performed at
the one-loop level,  it is well known that such relations 
between  $\Lambda$'s defined in different renormalization schemes are
actually  one-loop exact  \Cel, regardless of the presence of supersymmetry.
(In contrast,
the definition \LAMdef\ of the dynamical scale itself is one-loop
exact under the RG of the Wilsonian effective action, but only because of 
supersymmetry \refs{\SVone, \Dine}.)

Next we consider the cases $0<N_F<4.$ Let us introduce the symmetric
polynomials in the masses:
\eqn\inv{M^{(N_{F})}_{0} =  1\,,\quad
M^{(N_{F})}_{1} = \sum_{i=1}^{N_{F}} m_{i}^{2}\,,\quad
M^{(N_{F})}_{2} = \sum_{i<j}^{N_{F}} m^{2}_{i}m^{2}_{j}\ \ ,\ \ \cdots\ \ ,\ \
M^{(N_{F})}_{N_{F}}= \prod_{j=1}^{N_{F}} m_{j}^{2}\,.}
The  curves are then given by \SWtwo\foot{A technical aside: In the
absence of quark masses we have identified alternative curves for both
$N_F=2$ and $N_F=3$ with the desired singularity structure. They are
$y^2=x^2(x-u)-\textstyle{{9\over64}(\tL_2)^4(x-u/9)}$ and
$\textstyle{y^2=x^2(x-u)-{1\over64}(\tL_3)^2u^2}$, respectively.
Consistent with the physical arguments in \SWtwo, in the former case
the discriminant $\Delta(u)$ correctly factors into a product of two
double roots, while in the latter case $\Delta(u)$ is the product of
a simple root with a quartic root. However, in each case $\Delta(u)$ has
the wrong singularity structure once mass terms are added, which can be
seen most easily in the case that all masses are set equal.}
\eqn\curves{
y^{2}_{\sst (N_{F})} \ =\  
x^{2}(x-\tilde u)\,+
\,{1\over 4}\sqrt{M^{(N_{F})}_{N_{F}}}\,(\tL_{N_{F}})^{4-N_{F}}\,x
\,-\,{1\over 64}(\tL_{N_{F}})^{8-2N_{F}}
\sum_{\delta=0}^{N_{F}-1}M^{(N_{F})}_{\delta}
(x-\tilde u)^{N_{F}-1-\delta}}
Seiberg and Witten simply
equate $\tilde u\equiv u$ for $N_F=1,2,3$ but, as reviewed
earlier, for $N_F=3$ this should be corrected to 
 Eq.~\unought. It is easily checked that in the
decoupling limit $m_{N_F}\rightarrow\infty,$ one obtains the desired
result 
$ y^{2}_{\sst (N_{F})} \rightarrow y^{2}_{\sst (N_{F}-1)}$ if and
only if one makes the identification
\eqn\mrelated{m_{\sst N_{F}}\cdot (\tL_{N_{F}})^{4-N_{F}}\  =\
(\tL_{N_{F}-1})^{4-(N_{F}-1)}\ .}
Notice that this
is the same recursion relation as for the PV and DRED schemes, 
Eq.~\Lamrecursion. From the $N_F=0$ ``boundary condition'' \pvsw,
one immediately derives the dictionary between schemes given
in Eq.~\relrel\ above.

Before passing to the case $N_F=4$, we  note that the perturbative,
one-loop structure of the effective $U(1)$ coupling contained in 
Eq.~\newdkmmat\ has rarely appeared correctly in the literature.
This expression has recently been confirmed by \DHone;
rather than invoke the Weinberg prescription, these authors extract
the result directly from the elliptic curves, for arbitrary
$N_c.$

\newsec{Resolving the discrepancies in the $N_F=4$ solution}

We now turn to the interesting case  $N_{F}=4$. In this model
the $\beta$-function vanishes and  no dynamical scale is generated.
Note that it is trivial to extend the RG matching relation \Lamrecursion\ to
this case. The relation \nothresh\ still holds when $\mu=m_4
\rightarrow\infty$; consequently
\eqn\newmatch{m_4\,\exp(-8\pi^2/g_4^2)\ =\ \Lambda_3}
using Eq.~\LAMdef. Here, and henceforth, $g_4(\mu)\equiv g_4$ is the
scale-independent microscopic coupling in the PV or DRED scheme\foot{One
may ask why, in this finite theory, it is nevertheless necessary to
specify a scheme. As mentioned earlier, 
this is because the finiteness is due to an
``$\infinity$ minus $\infinity$'' cancellation between divergent
diagrams, which is intrinsically ill defined.
For this reason we cannot simply construct physical quantities directly
from the bare tree-level coupling $g_B$, as would be natural to do if all
individual graphs converged. That $g_B$ cannot be a scheme-independent
physical quantity can be seen from Eq.~\bare, which applies equally
to the DRED and DREG schemes, and which would then imply for $N_F=4$:
$g_\DRED(\mu)=g_\DREG(\mu)=g_B\,\mu^{2-D/2}$ in contradiction to
Eq.~\schemes.};  it combines with $\theta_4$ to form a single 
scale-independent PV or DRED holomorphic parameter
\eqn\taudef{\tau  \ = \ {8 \pi i \over g_4^2 } + 
   {\theta_4  \over  \pi} \ .}

Before turning to the elliptic curve, let us revisit the three discrepancies
with instanton physics that we wish to address. The first is the
discrepancy between Eqs.~\exactness\ and \tauresult\ which
 relate  $\taueff^\zero$ and $\tau.$
(As above, we will use the superscript $\scriptstyle(0)$ to denote the 
massless case.) The one-loop constant $c_0$ in Eq.~\tauresult\ was
not considered in Ref.~\dkmfour. In light of the preceding discussion,
we can now read off the value $c_0=4\log2$ from Eq.~\newdkmmat,
which is easily extended to the case $N_F=4$:
\eqn\newestdkmmat{ \taueff\ \equiv\ \hf\F''(a)\ = \ \tau \ + \ 
 {4i \over  \pi}\log 2 \ - \ 
{i \over 2 \pi} \sum_{f=1}^{4}
\ \log\,\big(\,1-m_f^2/2a^2\,\big)\ +\ {\cal O}(q)\ .}
An independent derivation of this 1-loop relation (which tests our
proposed reinterpretation of the massive curve) is discussed below.

The second and third discrepancies involve properties of the massive
$N_F=4$ curve (Sec.~16.3 of \SWtwo). Consider the RG decoupling property when
 the quark mass $m_4$ grows large. In the double scaling limit
the curve indeed collapses to the $N_F=3$ curve \curves; however,
as reviewed in Appendix B, this limiting behavior requires the 
identification \SWtwo
\eqn\newid{64\,m_4\exp\big(-8\pi^2/(g^\sw_4)^2\big)\ =\ \LSW_3 \ =\ 4\Lambda_3}
where the second equality follows from Eq.~\relrel. 
This apparently contradicts Eq.~\newmatch\ above; more precisely
it means that the coupling $g_4^\sw$ used in \SWtwo\
cannot be the expected DRED quantity $g_4.$
And thirdly, the $N_F=3$ curve that one flows to in this way has
$\tilde u=u$ rather than the shifted definition \unought.

As the reader can anticipate, we will posit that the parameter
$\tau$ that appears pervasively in the Seiberg-Witten curve for $N_F=4$ should
really be identified with the effective massless $U(1)$  coupling
$\taueff^\zero$, Eq.~\tauresult, rather than with
the microscopic $SU(2)$ coupling $\tau$, Eq.~\taudef. 
By definition, this reinterpretation
resolves the first of the three discrepancies. Pleasingly
it also resolves the second discrepancy: the constant factor
$c_0=4\log2$, when exponentiated,
precisely compensates for the factor of 16 mismatch
between Eqs.~\newid\ and \newmatch.
(A more stringent test of our proposal is discussed below.)
 Finally, the third discrepancy
will be resolved by altering the relation proposed in \SWtwo\ between
$\tilde u$ and $u$; however, there is an infinite ambiguity in
this procedure that we do not know how to eliminate.

We now review the Seiberg-Witten curve (Sec.~16 of \SWtwo), starting
with the massless case:
\def\tausw{\tau_\sw}
\eqn\nomass{\big(y^\zero\big)^2\ =\ x^3-\quarter g_2(\tausw)\,x\tilde u^2
-\quarter g_3(\tausw)\,\tilde u^3\ =\ W_1^\zero W_2^\zero W_3^\zero\ ,
\quad W_i^\zero=x-e_i(\tausw)\,\tilde u\ .}
Here $g_2$ and $g_3$ are rescaled Eisenstein series. The cubic roots
$e_i$ may be defined in terms of $\theta$-functions; they have the
semiclassical expansion
\eqn\expdef{
e_1 (\tau)=
\textstyle{2 \over 3} + 16q^2+16q^4+\CO(q^6) \ ,\quad
e_2 (\tau)=
-\textstyle{1 \over 3} - 8 q -8q^2-32q^{3} -8q^4
+\CO(q^{5})\ ,}
with $e_3=-e_1-e_2$.
As noted in \SWtwo, the $e_i$ are not strictly speaking modular forms
of $SL(2,\bigZ)$. Rather, they are weight-two modular forms of three
different conjugate subgroups of $SL(2,\bigZ)$; under the action of
the full group they permute amongst themselves. The curve \nomass\ is
well known in the math literature \Koblitz. It is designed so that if
the VEVs $a$ and $a_D$ are extracted in the standard way as periods
of the curve \SWtwo,
\eqn\periods{{da\over d\tilde u}\ =\ {\sqrt{2}\over8\pi}\int_{\gamma_1}
\,{dx\over y^\zero}\ ,\qquad
{da_D\over d\tilde u}\ =\ {\sqrt{2}\over8\pi}\int_{\gamma_2}
\,{dx\over y^\zero}\ ,}
then one has simply
\eqn\veryeasy{a\ =\ \sqrt{\tilde u/2}\ ,\qquad a_D\ =\ \tausw\,a\ .}
As expected, these are the defining equations of a classical free field theory,
with $\F^\zero(a)=\tausw\,a^2$ and $\tilde u=2a^2$. 

Seiberg and Witten 
make the two further assumptions $\tausw=\tau$ and $\tilde u=u$. 
The first of these assumptions contradicts Eq.~\tauresult;
instead, we will  assume
$\tausw=\tau_\eff^\zero$ as explained above. 
As pointed out in \HS,
this, in turn, invalidates the second assumption as well; instead,
one must take 
\eqn\uti{\tilde u\ =\ u\cdot \big(d\tau_\eff^\zero/d\tau\big)^{-1}\ . } 
This latter
redefinition is specific to the massless model; it
follows directly from the instanton version \refs{\Fucito,\dkmtwo,\HS}
of Matone's relation, which for four massless flavors reads,
\eqn\insmat{ u \ =  2\pi i q {\partial \F^\zero \over \partial q} \ ,}
combined with Eq.~\veryeasy.

We turn finally to the massive curve. Setting $e_{ij}=e_i-e_j$, one has \SWtwo:
\eqn\mostgeneralf{y^2=W_1W_2W_3
+ e_{12}e_{23}e_{31}\left(W_1T_1e_{23}+W_2T_2e_{31}+W_3T_3e_{12} \right)
- e^2_{12}e^2_{23}e^2_{31} N\ .}
Here 
$W_i=W_i^\zero- e_i^2\,R$,
where $R,$ $N$ and $T_i$ are  symmetric polynomials in the four masses:
\eqn\masspoly{\eqalign{R\ &=\ {1\over2}\sum_im_i^2\ ,
\qquad N\ =\ {3 \over 16}\sum_{i>j>k}m_i^2m_j^2m_k^2 -
{1 \over 96}\sum_{i \not= j} m_i^2m_j^4 + {1 \over 96}\sum_i m_i^6 \ ,\cr
T_1\ &=\ {1 \over 12}\sum _{i>j}m_i^2m_j^2 - {1\over 24}\sum_i m_i^4 \ ,
\quad
T_2\ =\ -{1 \over 2}\prod_i m_i - {1 \over 24}\sum _{i>j}m_i^2m_j^2 +
{1\over 48}\sum_i m_i^4 }}
with $T_3=-T_1-T_2.$ Seiberg and Witten observe that the coefficients of this
curve are $SL(2,\bigZ)$ modular invariants, provided that the 
aforementioned
 permutations amongst the $e_i$ are accompanied by the same permutations
acting on the $T_i$ (this is referred to as $SO(8)$ triality). 
They view this as strong circumstantial evidence
that the dyon spectrum itself is $SL(2,\bigZ)$ invariant.

 In restoring agreement with instanton calculations we need not
tamper with these important
$SL(2,\bigZ)$ properties. Rather, we will reinterpret
the intrinsic parameters of the curve. In particular we will continue to
take $e_i\equiv e_i(\tau_\eff^\zero)$ where $\tau_\eff^\zero$ is given by
the instanton series \tauresult. It remains only to relate the
parameter $\tilde u$ (which enters through the $W_i$) to the physical
quantum modulus $u=\langle\Tr\uA^2\rangle$. Dimensional analysis,
$O(4)$ symmetry, smoothness in the masses,
and the above-given massless limit suggest the following generic relation:
\eqn\implied{\tilde u\ =\ u\cdot\big({d\tau_\eff^\zero\over d\tau}\big)^{-1}
\ +\
R\cdot\sum_{n=0,2,4\cdots}\alpha_n\,q^n\ ,\quad 
q=\exp(i\pi\tau)\ .}
(This expression
is to be compared with the Seiberg-Witten proposal \SWtwo
\eqn\uudef{\tilde u\ =\ u\ -\ \hf e_1(\tau) R}
which is already faulty in the massless limit.) The absence of
odd instanton contributions is due to the discrete $\bigZ_2$ symmetry
discussed earlier, the mass parameter $R$ being even under this symmetry.
The numerical coefficients $\alpha_n$ in \implied\  may be constrained
using a variety of physics considerations explained in Appendix B.
We find that  $\alpha_0
=-1/3,$ and $\alpha_2\,=\,37/(3^3\,2^5)\,$; the latter value disagrees again
with the proposed relation \uudef, but is necessary to recapture the
shifted definition \unought\ of $\tilde u$ in the decoupling limit.

We conclude with the following comments:

\bf(i) \rm
The higher-instanton contributions $\alpha_n$ with $n=4,6,\cdots$
remain completely undetermined, as do the constants $c_n$ with $n=4,6,\cdots$
in the relation \tauresult\ between $\taueff^\zero$ and $\tau.$ When
masses are incorporated into Eq.~\tauresult, Matone's relation
\refs{\MAtone,\BMSTY,\dkmtwo} may give an interesting correspondence between
the two series (as it already does in the massless \hbox{case \HS).} 

\bf(ii) \rm
Our lack of complete knowledge of the relation between $\tilde u$ and $u$
 does not
actually affect the low-energy effective $U(1)$ Lagrangian. This is
because
 $a$ and $a_D$ are still determined by Eqs.~\periods\ (with $y$ instead of 
$y^\zero$ in the massive case). Both sides of these equalities 
involve the (independent)
 variable $\tilde u$ rather than the unknown
(dependent) variable $u$. The former is then eliminated in
favor of $a$, giving a prepotential in which neither $u$ nor $\tilde u$
appears: $\F\,=\,\F\big(a,\tilde u(a);\{m_i\}\big)\,\equiv\,\F(a;\{m_i\}).$
In other words, so long as the classical VEV $a$ is considered
the independent variable (and not the quantum modulus $u=\langle\Tr\uPhi^2
\rangle$ on which
$a$ depends in a presently unknown way, and \it vice versa\rm), 
the prepotential is ``known'' as a function of $\taueff^\zero$ (albeit
not as a function of the microscopic $\tau$). 
In contrast, for $N_F<4,$ both $a$ and 
$\F$ are known functions of $u$ as well.

\bf(iii) \rm
As a stringent test of our proposed redefinitions $e_i\equiv
e_i(\taueff^\zero)$, we have in fact constructed $\F(a)$ as outlined
in \bf (ii)\rm, and verified that the right-hand side of 
Eq.~\newestdkmmat\ is indeed reproduced by the curve (paralleling
Ref.~\DHone).

\bf(iv) \rm
Finally we reiterate the point that the massive $N_F=4$ curve is
still $SL(2,\bigZ)$ invariant, but only in terms of the rather non-intuitive
quantity $\taueff^\zero$ (the effective coupling in the \it massless \rm
theory) 
rather than the microscopic coupling $\tau$. While the relation 
\tauresult\ between these parameters is currently unknown beyond the
2-instanton level, it would be pleasing if, in the end, the model
turned out to be modular invariant in terms of $\tau$ as well.

$$\scriptscriptstyle ***************************$$

We acknowledge useful discussions with
T. Bhattacharya, V.A. Khoze and  M.J. Slater.
The work of ND was supported in part by a PPARC Advanced Research
Fellowship; both ND and
 VVK were supported in part by the Nuffield 
Foundation; MM was supported by the Department of Energy.

\appendix{A}{Note on conventions}

Here we provide the dictionary between the normalization conventions 
of Ref.~\SWtwo\ which we have adopted in this paper, and those
of  Ref.~\SWone\ which we used in previous work \refs{\dkmone,\dkmfour}.

The original  unrescaled VEV definitions of \SWone\ are:
$$\langle\uA\rangle\ =\ \vhiggs\tau^3/2\ ,\quad
\langle\uA_D\rangle\ =\ \vhiggs_D\tau^3/2\ ,\quad
 \vhiggsd \ = \ {\partial \F (\vhiggs) \over \partial \vhiggs}\ .$$
$$ \taueff \ \equiv \ {8 \pi i \over \geff^2 } + 
   {\thetaeff  \over  \pi} \ = \  
 2 {\partial \vhiggsd \over \partial \vhiggs} 
  \ = \ 2 {\partial^2 \F (\vhiggs) \over \partial \vhiggs^2} \ . $$
The  VEV normalizations of \SWtwo\ adopted in this paper are:
$$ a \ = \ \hf \vhiggs \ , \ \ \ \ 
   a_D \ = \ \vhiggsd \ = \ 
   \hf {\partial \F (a) \over \partial a}  $$
$$ \taueff  \ = \  
  {\partial a_D \over \partial a} 
  \ = \ \hf {\partial^2 \F (a) \over \partial a^2}  $$    
$$ \vhiggs \ = \ \sqrt{2u} + \cdots \ , \ \ \ \ 
    a \ = \ \hf \sqrt{2u} + \cdots  $$

The prepotential for  $N_F < 4$ massless flavors
in the two different normalizations reads:
$$
{\cal F}^{(N_{F})}(a, \tL_{N_F})
 \ =\ {\cal F}^{(N_{F})}_{\rm pert}
- {i \over \pi}\sum_{n=2,4,6,\ldots}  F^{(N_{F})}_{n}\left(
{\tL_{N_{F}} \over a}\right)^{n(4-N_{F})}a^{2} \, $$
$$
{\cal F}^{(N_{F})}(\vhiggs, \Lambda_{N_F})
 \ =\ {\cal F}^{(N_{F})}_{\rm pert}
- {i \over \pi}\sum_{n=2,4,6,\ldots}  \F^{(N_{F})}_{n}\left(
{\Lambda_{N_{F}} \over \vhiggs}\right)^{n(4-N_{F})}\vhiggs^{2} \ . $$
Using Eq.~\relrel, the relation between the instanton coefficients
is determined to be:
$$ {\cal F}^{(N_{F})}_{n} \ 
= \ 2^{n(6-N_{F}) -2} \ F^{(N_{F})}_{n} \ .$$

\appendix{B}{Constraints on the relation between $\tilde u$ and $u$}

In this appendix we constrain some of the \it a priori \rm unknown numerical
constants $\alpha_n$ that relate $\tilde u$ to $u$ as per Eq.~\implied.
We use the following two considerations:


\bf(i) \rm
Closely 
following Sec.~16.3 of \SWtwo, we first consider the illuminating special
case $(m_1,m_2,m_3,m_4)=(0,0,m,m).$ Then $N=T_i=0$ and one has simply
$y^2=W_1W_2W_3$ with $W_i=x-e_i\tilde u-e_i^2m^2.$ As always, 
the critical points on the quantum moduli space are the values of $u$
for which two of the three $x$ roots coincide. In the present instance
this means $e_i\tilde u+e_i^2m^2=e_j\tilde u+e_j^2m^2,$ or 
equivalently since $\sum e_i=0\,:$ 
\eqn\appa{\textstyle{\tilde u\ =\ \{e_1m^2,e_2m^2,e_3m^2\}\ \simeq\ 
\{{2\over3}m^2\,,\,(-{1\over3}-8q)m^2\,,\,
(-{1\over3}+8q)m^2\}}}
using Eq.~\expdef.
We have dropped terms of order $q^2$ as these do not survive the double
scaling limit. Now consider
the decoupling limit $m\rightarrow\infty.$ On the one hand, one expects
a perturbative singularity at $u\simeq 2a^2\simeq m^2$ which corresponds, 
physically, to two quark multiplets becoming massless (see 
Eq.~\newestdkmmat). On the other hand, the leftover model after
the two heavy flavors decouple is the massless $N_F=2$ theory, which
has singularities at $u=\pm\eighth (\tL_2)^2$ (see Eq.~\curves). In sum,
\eqn\appb{u\  \simeq\ 
\{m^2\,,\,-\eighth (\tL_2)^2\,,\,\eighth (\tL_2)^2\}\ .}
Equating \appb\ with \appa\ forces $\alpha_0=-1/3$; and
 furthermore $8\qeff^\zero\,m^2\,=\,\eighth (\tL_2)^2$ which is precisely
consistent with the DRED recursion relations \mrelated, \newmatch\ and \newid\
(with $g^\sw_4\equiv g_\eff^\zero$).

\bf(ii) \rm
Returning to the case of four generic masses, we next consider
the double scaling limit $m_4\rightarrow\infty,$ $\qeff^\zero\rightarrow0$
with the product $m_4\,\qeff^\zero$ fixed at $\tL_3/64$ as per
Eq.~\newid. Naively, we would like the curve \mostgeneralf\ to collapse
to the $N_F=3$ curve defined by Eqs.~\curves\ and \unought. Instead,
in this limit the right-hand side of \mostgeneralf\ diverges badly.
However we can exploit the fact that the variable $x$ is just a dummy of 
integration (see Eq.~\periods); an appropriate shift in $x$ eliminates
this divergence and guarantees a smooth RG limit. Accordingly we let
\eqn\xshift{\textstyle x\ \longrightarrow\ x\ +\ \big(\,{2\over9}
+({1\over32}-{1\over3}\alpha_2)\,q^2+\CO(q^4)\,\big)\,R\ -\ \big({1\over3}+
\CO(q^2)\big)\,u}
which obscures the $SL(2,\bigZ)$ properties of the curve, but preserves
the $\bigZ_2$ properties. 
The explicit factors in \xshift\ have the following genesis: the constant $2/9$
eliminates the large-mass
divergence; the constant $1/3$ guarantees that the cubic
terms in $x$ and $u$ will have precisely
the form $x^2(x-u)$ dictated by Eq.~\curves;
and finally the factor $(1/32-\alpha_2/3)$ eliminates the $(\tL_3)^3\,
m_1m_2m_3$ term that otherwise generically appears, again in order to harmonize
with Eq.~\curves. (This term is not forbidden by any symmetry; its absence
from the $N_F=3$ curve \curves\ is convention dependent, as is the form
of the cubic term.)
Substituting Eq.~\xshift\ into the $N_F=4$ curve
\mostgeneralf\ and taking the double scaling limit then indeed reproduces
the $N_F=3$ curve \curves, provided that
\eqn\alphatwo{-(1+32\alpha_2)\,2^{-10}\ =\ u_0\ .}
{}From the value of $u_0$ quoted in Eq.~\unought\ one deduces
$\alpha_2\,=\,37/(3^3\,2^5)\,.$

\listrefs
\bye